\documentclass[journal]{IEEEtran}
\usepackage{ifpdf}

\usepackage{cite}

\ifCLASSINFOpdf
\usepackage[pdftex]{graphicx}
\else
\fi
\usepackage{amsmath}
\usepackage{array}

\usepackage{stfloats}
\usepackage{url}

\usepackage[ruled]{algorithm2e}
\usepackage{algorithmic}

\usepackage[caption=false,font=footnotesize]{subfig}

\begin{document}
%
\title{Smart Contract-based Spectrum Sharing Transactions for Multi-Operators Wireless Communication Networks}
%
%
%

\author{Shuang~Zheng,
       Tao~Han,~\IEEEmembership{Member,~IEEE}
       , Yuna~Jiang,
       and~Xiaohu~Ge, ~\IEEEmembership{Senior~Member,~IEEE}
}

\maketitle

\begin{abstract}
Multiple-operators (multi-OPs) spectrum sharing mechanism can effectively improve the spectrum utilization in fifth-generation (5G) wireless communication networks. The secondary users are introduced to opportunistically access the licensed spectrum of idle operators (OPs). However, the identity privacy and data security issues raise great concerns about the secure spectrum sharing among multi-OPs. To address these challenges, a permissioned blockchain trust framework is proposed for the spectrum sharing in multi-OPs wireless communication networks in this paper. The Multi-OPs Spectrum Sharing (MOSS) smart contract is designed on the constructed permissioned blockchain to implement the spectrum trading among multi-OPs. Without the need of trustless spectrum broker, the MOSS smart contract enforces multi-OPs to share the spectrum truthfully and designs a punishment mechanism to punish malicious OPs. The MOSS smart contract is tested on the Remix integrated development environment (IDE) and the gas costs of MOSS smart contract is estimated. The performance analysis of the proposed permissioned blockchain trust framework demonstrates that the privacy, openness and fairness of the proposed solution are better than traditional spectrum allocation solutions.
\end{abstract}

\begin{IEEEkeywords}
Spectrum sharing, permissioned blockchain, smart contract, privacy, security.
\end{IEEEkeywords}

%
\IEEEpeerreviewmaketitle

\section{Introduction}
%
%
%
%
\IEEEPARstart{W}{ireless} spectrum, as a scarce natural resource, is an essential foundation to support wireless communications. The exclusive spectrum is staticlly allocated to an operator (OP) in traditional spectrum management solutions. Based on the statistics in the Federal Communications Commission's investigation report \cite{FCC1}, the utilization of the licensed spectrum ranges from 15\% to 85\% with traditional static spectrum allocation solutions. The communication traffic has exploded with the rise of high definition (HD) video transmission, virtual reality (VR), augmented reality (AR) and other high-bandwidth services in fifth-generation (5G) wireless communication networks \cite{cisco2}. The traditional static spectrum allocation and management solutions  would cause the spectrum underutilization and inefficiency, which have great limitations on meeting traffic needs of 5G wireless communication networks. Therefore, it is urgent to find new solutions to improve the spectrum utilization for the spectrum management and allocation in wireless communication networks.

Dynamic spectrum sharing in cognitive radio (CR) networks is a feasible solution to relieve the shortage and underutilization of spectrum resources \cite{molita3},\cite{haykin4},\cite{koufos5}. CR is defined as an intelligent wireless communication system which can sense and learn from the surrounding environment by the understanding-by-building method \cite{haykin4}. Based on the cognitive and reconfigurable abilities of CR, primary users (PUs) can share its licensed spectrum with secondary users (SUs) to improve the spectrum utilization \cite{if6}. The microeconomic theories, such as the auction theory \cite{wang12}, the game theory \cite{yu13}, the contract theory \cite{gao18},\cite{liang19} and the price theory \cite{hossian21}, were adopted to realize the dynamic spectrum sharing between PUs and SUs. Two main architectures have been proposed in previous studies of dynamic spectrum sharing among multi-operators (multi-OPs): the centralized \cite{behera22} and the distributed \cite{wang23} spectrum sharing architectures. Although both architectures can relieve the spectrum shortage and improve the spectrum utilization, security issues of interaction among mutual-trustless entities are severe and have not been considered in previous studies. In the centralized spectrum sharing architecture, a spectrum broker is introduced to be responsible for the spectrum allocation between participating PUs and SUs. In order to share the spectrum efficiently, both PUs and SUs need to interact the private information with the trustless spectrum broker \cite{lin24}. In this case, the interaction suffers from a single-point failure and privacy disclosure. In the distributed spectrum sharing architecture, participating PUs and SUs need to interact with each other \cite{gel25}, which may face the challenge of huge communication cost and privacy leakage. Therefore, it is necessary to propose an efficient solution to guarantee the secure interaction in the dynamic spectrum sharing among multi-OPs.

Recently, the promising blockchain \cite{naka26} and smart contract \cite{chris27} technologies have been widely used in many fields due to its decentralization and security, such as the medical treatment \cite{jiang29}, the big data \cite{zhang30}, the electrical vehicles \cite{wang31}, and the Internet of things (IoT) \cite{zhang32}, etc. The smart contract consists of many predefined functions which can be triggered by transactions to realize specific functionalities. Transactions are recorded on the distributed ledger maintained by blockchain consensus nodes. The blockchain has also been applied to the spectrum sharing with advantages of security, fair and transparancy. The ownership and usage of spectrum are recorded on the distributed ledger of blockchain \cite{bayhan33},\cite{zubow34}. The security and privacy-preserving of spectrum sharing among multi-OPs can be guaranteed without the existence of a trustless spectrum broker \cite{weiss35},\cite{bilen37}. However, due to the high block verification delay and low throughput of the public blockchain, the efficient spectrum sharing cannot be achieved in existing blockchain studies.

To overcome these issues, in this paper, a permissioned blockchain is constructed for the spectrum sharing in multi-OPs wireless communication networks. The practical byzantine fault tolerant (PBFT) consensus algorithm instead of proof of work (POW) is adopted in the permissioned blockchain, which can greatly realize the the high throughput and short delay. The multi-OPs acted as authorized nodes are required to be authenticated the certificate by the administrator before joining in the constructed permissioned blockchain. In this paper, the Multi-OPs Spectrum Sharing (MOSS) smart contract is designed on the blockchain for spectrum sharing in wireless communication networks. Without a trustless spectrum broker, different OPs can autonomously trade the spectrum by calling functions defined in the MOSS smart contract.

The main contributions in this paper are described as follows.
\begin{enumerate}
\item Based on the constructed permissioned blockchain, a decentralized and secure framework is proposed for the spectrum sharing in multi-OPs wireless communication networks.
\item The double auction and free-trading market are introduced to design the MOSS smart contract, which enables multi-OPs to autonomously share the spectrum in wireless communication networks.
\item The MOSS smart contract is compiled on the Remix integrated development environment (IDE). The gas cost of each function defined in the MOSS smart contract is estimated. The performance analysis of the proposed spectrum sharing solution are given.
\end{enumerate}

The rest of this paper is organized as follows. In Section II, related works are firstly discussed. In Section III, a permissioned blockchain trust framework is proposed for the spectrum sharing in multi-OPs wireless communication networks. In Section IV, the MOSS smart contract is designed. In Section V, experimental results are presented. The performance analysis are also discussed. Finally, Section VI concludes the paper.

\section{RELATED WORK}
In this section, related works about the spectrum sharing with the CR technology are discussed, and the existing studies of the spectrum sharing with the blockchain technology are also introduced.

Many studies combine the CR technology with the microeconomic model to solve the challenges of spectrum shortage and underutilization. The auction-based approach was designed in \cite{wang12} to solve the spectrum sharing between macro base stations (MBS) and femto access points (FAP), who acted as the bidder for the additional spectrum of MBSs.  The many-to-one stable matching game theory and the stochastic geometrical approaches were used in \cite{yu13} to tackle the dynamic spectrum sharing among network OPs. The contract and price theory were also applied to the spectrum sharing in CR networks. Considering the scenario consisted of a single PU and multi-SUs, an optimal contract was derived to satisfy the incentive compatibility and individual rationality \cite{gao18}. A hybrid distributed-centralized spectrum sharing framework was proposed in \cite{liang19} based on the contract theory, where a random leader mechanism was introduced to randomly elect the spectrum manager among SUs or PUs. By using the Equilibrium price theory , a distributed algorithm based on the Bertrand game and repetitive game was proposed to maximize the benefits of multi-PUs and satisfy the demands of SUs \cite{hossian21}. However, these studies did not consider the security issue in the spectrum sharing among multi-OPs, which may cause the privacy disclosure of OPs.

The emergence of blockchain technology provides a new way to ensure the security and decentralization. There exist some studies applying the blockchain and smart contract for various applications. A healthcare information exchange blockchain platform was built in \cite{jiang29}, which combined the off-chain storage and on-chain verification to ensure the privacy and authentication of healthcare data. An energy blockchain was proposed to provide secure charging services for electric vehicles \cite{wang31}. An IoT E-business model was proposed in \cite{zhang32}, where data of transactions were recorded on the blockchain. However, only a few studies have applied the blockchain to the scenario of spectrum sharing in wireless communication networks. A Spectrum Sensing as a Service (SPASS) architecture was proposed to realize the payment transferring of outsourced spectrum sensing services through the smart contract \cite{zubow34}. The feasibility of applying the smart contract-enabled blockchain technology to the spectrum sharing was analyzed in \cite{weiss35} from five aspects, including the decentralization, the transparency, the irreversibility, the availability and the security. In \cite{pas39}, considering that some home or business users could receive a certain revenue by providing spectrum sensing services for nodes in SU networks, a smart contract was designed to implement Service Level Agreements (SLAs) between mobile OPs and community service providers. However, the efficient spectrum sharing based on the blockchain is still an open issue due to the high block verification delay and low throughput.

\section{SYSTEM FRAMEWORK}
In this section, the system framework based on the permissioned blockchain for the spectrum sharing among multi-OPs in wireless communication networks is shown in Figure 1. The MBS and small cell base stations (SBS) of multi-OPs coexist in the specific region, where each OP includes a MBS covering a number of SBSs. The MBS and SBSs of each OP analyze the spectrum usage and user locations reported by user terminals in the specific region. The MBS and SBSs of each OP report analysis results to the corresponding servicer, i.e., the Operation Administration and Maintenance (OAM) servicer, which is responsible for the spectrum management among associated MBS and SBSs. Based on the analysis results collected from the MBS and SBSs, the OAM servicer of each OP can reasonably predict the spectrum demand during the period of $\left[ {{\text{t}}_{\text{b}}}\text{ , }{{\text{t}}_{\text{e}}} \right]$.
\subsection{MAIN ENTITIES OF THE SPECTRUM SHARING NETWORK}
A spectrum trading market can be established to satisfy different spectrum demands of multi-OPs. The MOSS smart contract is designed to realize the spectrum trading and allocation among multi-OPs. The proposed permissioned blockchain trust framework consists of three types of entities for the spectrum sharing in multi-OPs wireless communication networks: seller OPs group, buyer OPs group and the administrator. These three entities are connected and communicated through the constructed blockchain networks.  The main entities of the proposed Spectrum Sharing Framework are listed as follows.

\begin{figure}[!h]
\centering
  \includegraphics[width=0.48\textwidth]{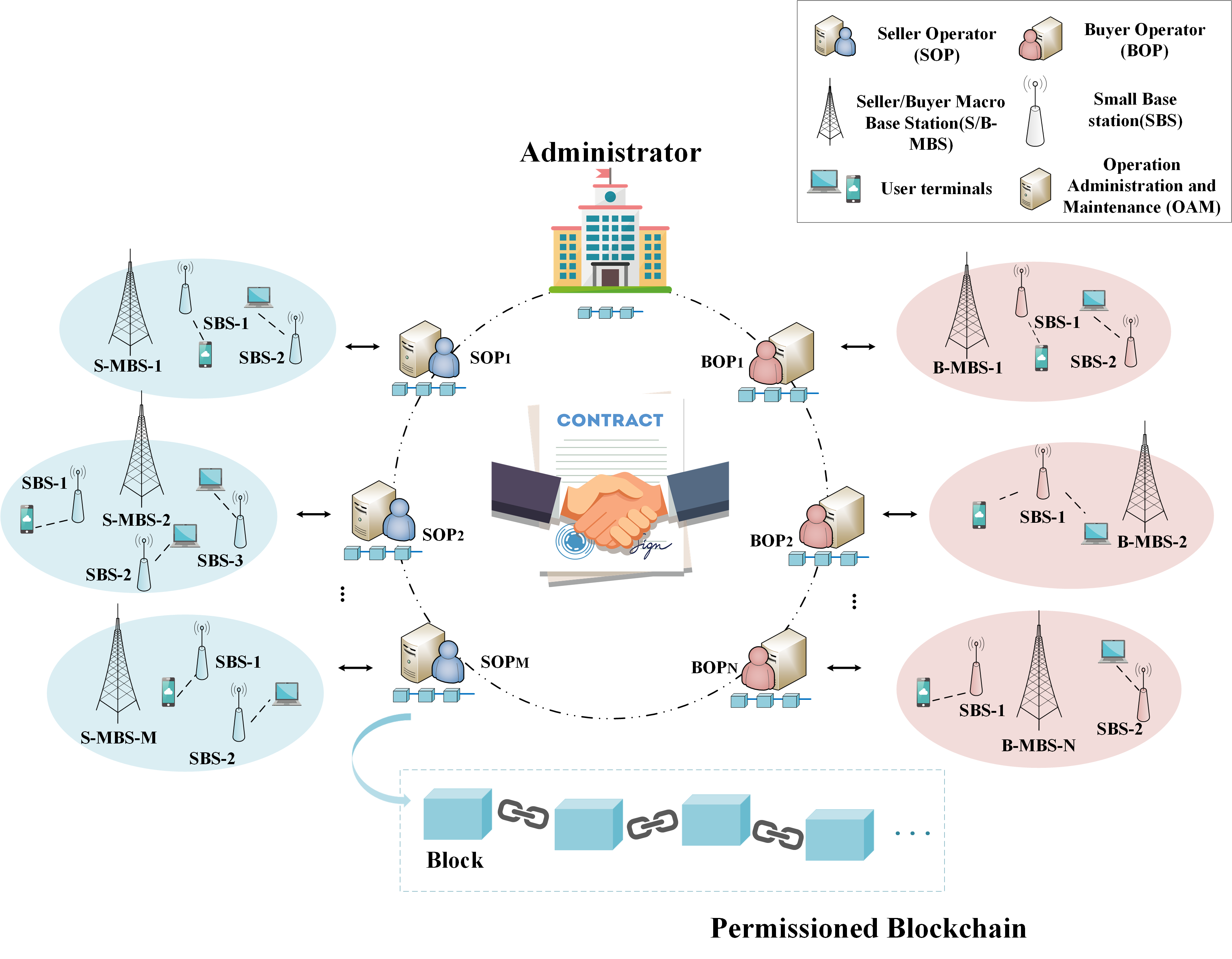}
\caption{System framework}
\label{Fig1}       
\vspace{-0.3cm}
\end{figure}

Seller OPs group: The seller OPs group is constituted by $\mathcal{M}$($\mathcal{M}\text{=}\left\{ 1,2,\cdots,M \right\}$)  OPs. The $m$-th OP decides to sell ${{B}_{m}}\left( {{B}_{m}}\le B_{m}^{tot},m\in \mathcal{M} \right)$ bandwidth, where $B_{m}^{tot}$ is the total bandwidth owned by the $m$-th OP. In order to guarantee the quality of service (QoS) of  service users associated with the $m$-th OP during the period of $\left[ {{\text{t}}_{\text{b}}}\text{ , }{{\text{t}}_{\text{e}}} \right]$, the $m$-th OP retains $B_{m}^{left}$ bandwidth to serve associated users. $B_{m}^{left}$ meets the following requirement:
\begin{equation}
B_{m}^{left}=B_{m}^{tot}-{{B}_{m}}\ge B_{m}^{req},\forall m\in \mathcal{M},
\label{eq}
\end{equation}
where $B_{m}^{req}$ is the bandwidth required by the $m$-th OP to provide the specific QoS for the associated users, ${{p}_{m}}$ is the price per unit bandwidth bid by the $m$-th OP.

Buyer OPs group: The buyer OPs group is constituted by $\mathcal{N}\left( \mathcal{N}=\left\{ 1,2,\cdots ,N \right\} \right)$ OPs. The $n$-th OP decides to buy ${{W}_{n}}$ bandwidth based on the prediction of spectrum demand during the period of $\left[ {{\text{t}}_{\text{b}}}\text{ , }{{\text{t}}_{\text{e}}} \right]$, and the bid price per unit bandwidth is set as ${{c}_{n}}$.

Administrator: The administrator is responsible for supervising the implementation process of spectrum sharing among multi-OPs. The main duties of the administrator are listed as follows:
\begin{enumerate}
  \item The administrator is responsible for the certificate authentication and the off-chain supervision of OPs. Each OP need to be authenticated the identity before acting as a node in the blockchain. Once supervising the malicious behaviour of OPs, the administrator can record it on the blockchain and punish malicious OPs according to the pre-defined rules in the smart contract.
  \item Considering the fact that the deployment of smart contract needs to cost large gas and selfish OPs are unwilling to pay for it, the administrator can serve as the deployer of the MOSS smart contract and the caller of some specified functions. What to mention is that the administrator just acts as the deployer and caller without participating in the process of spectrum allocation as a spectrum broker.
\end{enumerate}
In the proposed framework, the MOSS smart contract is deployed at time ${{\text{t}}_{\text{0}}}$ by the administrator. Simultaneously, the bidding time is initialized as ${{\text{t}}_{\text{bid}}}$ and the free-trading time is initialized as ${{\text{t}}_{\text{free}}}$ when the MOSS smart contract is deployed. The free-trading market is triggered at time ${{\text{t}}_{\text{1}}}$(${{\text{t}}_{\text{1}}}>\left( {{\text{t}}_{\text{0}}}+{{\text{t}}_{\text{bid}}} \right)$) by the administrator. The valid bidding duration is $\left[ {{\text{t}}_{\text{0}}}\text{  ,  }{{\text{t}}_{\text{0}}}+{{\text{t}}_{\text{bid}}} \right]$ ($\left( {{\text{t}}_{\text{0}}}+{{\text{t}}_{\text{bid}}} \right)<{{\text{t}}_{\text{b}}}$), and the valid free-trading duration is $\left[ {{\text{t}}_{\text{1}}}\text{ , }{{\text{t}}_{\text{1}}}+{{\text{t}}_{\text{free}}} \right]$($\left( {{\text{t}}_{\text{1}}}+{{\text{t}}_{\text{free}}} \right)<{{\text{t}}_{\text{b}}}$). The more details will be introduced in Section IV.
\subsection{PERMISSIONED BLOCKCHAIN-ENABLED SPECTRUM SHARING OPERATIONS}
To guarantee the trusted spectrum sharing among multi-OPs in 5G wireless communication networks, a permissioned blockchain-enabled spectrum sharing framework is established. The general spectrum sharing operations for entities in the permissioned blockchain are described as follows.
\begin{enumerate}
\item  Initializing the sysytem: the permissioned blockchain is only available to the authenticated entities \cite{wood42}. Before joining in the permissioned blockchain, the OP $i$ $\left( i\in \mathcal{M}\cup \mathcal{N} \right)$ needs to register with the administrator to be authorized with the identity $\text{I}{{\text{D}}_{i}}$, the public key $\text{P}{{\text{K}}_{i}}$, the private key $\text{S}{{\text{K}}_{i}}$, the legitimate certificate $\text{Cer}{{\text{t}}_{i}}$ and the wallet address $\text{W}{{\text{A}}_{i}}$. The mapping list $\left\{ \text{I}{{\text{D}}_{i}}\text{,P}{{\text{K}}_{i}}\text{,}{{\text{Cert}}_{i}}\text{,W}{{\text{A}}_{i}} \right\}$ can be sent as a transaction by the OP $i$ to adjacent authorized OPs. All the transaction records are stored in the memory pool of OAM of authorized OPs. The OAM of each OP is considered to have a certain storage and computing power. Based on the stored certificate information of each registered OP, the authorized OPs can verify the identity of transaction sender.
\item Sending transactions: after initialization, the administrator can deploy the MOSS smart contract to start the spectrum sharing among multi-OPs. Multi-OPs can choose to become the seller OPs or buyer OPs based on the spectrum usage state. Seller OPs and buyer OPs can trade the spectrum by calling specific functions defined in the MOSS smart contract. More details about the MOSS smart contract will be given in the section IV.
\item Performing the consensus process: the PBFT consensus algorithm is used to reach a consensus about the spectrum allocation result among registered OPs. Multi-OPs who want to participate in the consensus process can become a replica in the PBFT algorithm, and one of OPs can be selected as the primary. A set of transactions are verified and packaged into the block by the primary. The primary broadcasts the block to all replicas for auditing. The audited results of each replica are announced among all replicas in the permissioned blockchain. By comparing the received audited results, each replica can send the feedback information to the primary. After reaching a consensus, the block data will be broadcasted by the primary to each entity in the permissioned blockchain and a newly verified block can be added to the blockchain.
\end{enumerate}

Underloaded OPs who may have the idle spectrum during the period of $\left[ {{\text{t}}_{\text{b}}}\text{ , }{{\text{t}}_{\text{e}}} \right]$ can join the seller OPs group and share their idle spectrum with a certain price. As a consequence, the revenue and spectrum utilization of underloaded OPs can be improved. Overloaded Ops, whose licensed spectrum is unable to meet needs of all users during the period of $\left[ {{\text{t}}_{\text{b}}}\text{ , }{{\text{t}}_{\text{e}}} \right]$, can join the buyer OPs group to purchase the spectrum from the seller OPs group. With the spectrum purchased from seller OPs, the QoS of users served by overloaded OPs can be improved.

Both the spectrum trading and payment transferring between seller OPs and buyer OPs are executed by the designed MOSS smart contract. Each entity has a unique account. Transactions are recorded in the constructed permissioned blockchain. During the period of $\left[ {{\text{t}}_{\text{0}}}\text{  ,  }{{\text{t}}_{\text{0}}}+{{\text{t}}_{\text{bid}}} \right]$, $\text{M}$ seller OPs and $\text{N}$ buyer OPs independently register and set the bid through the MOSS smart contract. As the registration period ends, the administrator invokes the MOSS smart contract to perform the spectrum auction among registered seller and buyer OPs. If there exist unsuccessful matches between buyer OPs and seller OPs, the administrator can invoke the MOSS smart contract to open the free-trading market. Unsuccessfully matched OPs can choose whether to enter the free-trading market during the period of $\left[ {{\text{t}}_{\text{1}}}\text{ , }{{\text{t}}_{\text{1}}}+{{\text{t}}_{\text{free}}} \right]$, and adjust their bids based on matched results of the spectrum auction stage. Finally, all OPs can invoke the MOSS smart contract to take back the rest of their deposit. A round of spectrum sharing among multi-Ops during the period of $\left[ {{\text{t}}_{\text{b}}}\text{ , }{{\text{t}}_{\text{e}}} \right]$ is finished.

\section{SMART CONTRACT DESIGN}
In this section, the flow of designed MOSS smart contract is introduced. The explanation of parameters are shown in Table 1. The smart contract is mainly consisted of functions and events \cite{luu43}. The functions are the executable code to implement the specified functionality, and the event can be used to notify the authenticated nodes of the state change in the blockchain \cite{hu44}.

\begin{table}
\caption{Explanation of parameters.}
\label{table}
\setlength{\tabcolsep}{3pt}
\begin{tabular}{|p{80pt}|p{150pt}|}
\hline
Notations&
Descriptions\\
\hline
${{\text{t}}_{0}}$&
Time to deploy the MOSS
\\
${{\text{t}}_{\text{bid}}}$&
Time of bidding registration
\\
${{\text{t}}_{\text{free}}}$&
Time of free-trading market
\\
${{\text{t}}_{1}}$&
Time to begin the free-trading market
\\
${{B}_{m}}$&
The bandwidth selled by the $m$-th op
\\
${{p}_{m}}$&
Price per unit bandwidth of the $m$-th op
\\
${{W}_{n}}$&
The bandwidth bought by the $n$-th op
\\
${{c}_{n}}$&
Price per unit bandwidth of the $n$-th op
\\
\emph{asks}&
The registration set of seller OPs
\\
\emph{bids}&
The registration set of buyer OPs
\\
deposit&
The account balance of OP
\\
$addr_{m}^{S}$&
address of the $m$-th op
\\
$addr_{n}^{B}$&
address of the $n$-th op
\\
${msg}.{sender}$&
The sending address of transaction
\\
$\text{ }\!\!\_\!\!$ \emph{addr}&
The address of OP in the free-trading market
\\
$\text{ }\!\!\_\!\!$ \emph{price}&
The latest price of OP in the free-trading market
\\
$\text{ }\!\!\_\!\!$ \emph{bandwidth}&
The latest bandwidth of OP in the free-trading market
\\
\emph{role}&
The seller/buyer OP
\\
\hline
\end{tabular}
\label{tab1}
\end{table}

In the MOSS smart contract, the multi-OPs spectrum sharing process is divided into three stages: the registration of OPs, the spectrum trading and the payment clearing. Functions defined in the MOSS smart contract should be agreed by all OPs before the MOSS smart contract is deployed in the permissioned blockchain. The administrator sends a transaction to deploy the MOSS smart contract at time ${{\text{t}}_{\text{0}}}$. The address of MOSS smart contract is known by all OPs once this transaction is packaged into the valid block. The timestamp in the blockchain can be used in the MOSS smart contract to judge the timeliness of transactions. The function call that is invalid in the current period will not be packaged into the block by miners. The detailed flow of MOSS smart contract is shown in Figure 2 (the source code of MOSS smart contract can be seen in the github at \url{https://github.com/cherry1124/MOSS_CONTRACT}) and is described as follows.

\begin{figure}[!h]
\centering
  \includegraphics[width=0.48\textwidth]{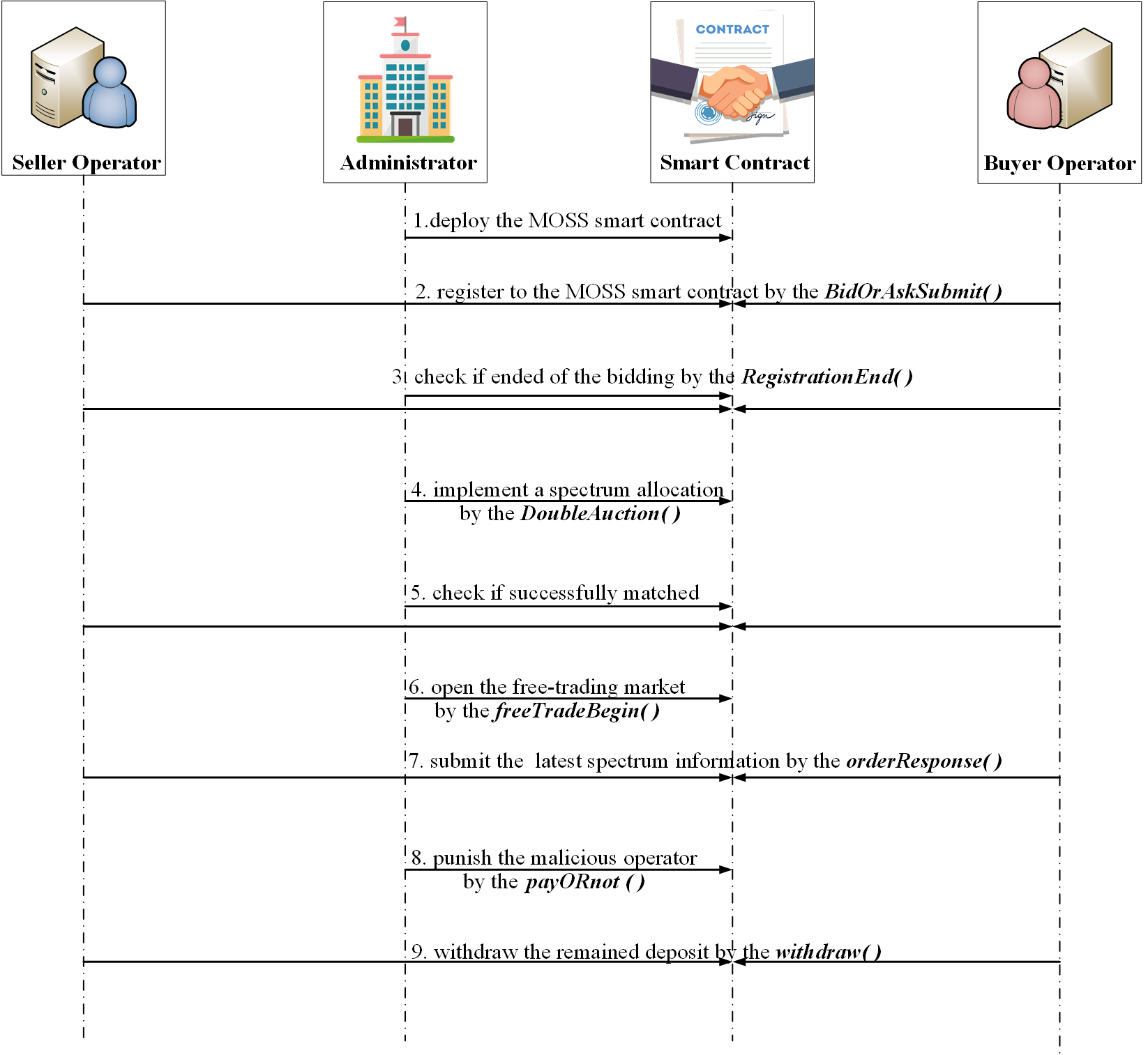}
\caption{Flow of the designed MOSS smart contract}
\label{Fig2}       
\vspace{-0.3cm}
\end{figure}

\subsection{The registration of OPs}
In the registration stage of OPs, the \textit{BidOrAskSubmit( )} function is designed to submit the bid of multi-OPs and the \textit{AuctionEnd( )} function is designed to judge whether the registration process is ended. The specific functionalities of \textit{BidOrAskSubmit( )} and \textit{AuctionEnd( )} are described as follows.

\textbf{\emph{BidOrAskSubmit( )}}: During the period of $\left[ {{\text{t}}_{\text{0}}}\text{  ,  }{{\text{t}}_{\text{0}}}+{{\text{t}}_{\text{bid}}} \right]$, OPs call the \textit{BidOrAskSubmit( )} when they want to participate in the spectrum sharing and declare whether they are buyer OPs or seller OPs. The bandwidth and price per unit bandwidth they want to buy or sell also need to be given. The \textit{BidOrAskSubmit( )} function is set as payable, which represents that the \textit{BidOrAskSubmit( )} have the functionality of ether transferring. All OPs must deposit more than 1 ether into the address of MOSS smart contract when the \textit{BidOrAskSubmit( )} is called. The ether is the common digital currency in the Ethereum. The requirement of deposit can protect the spectrum trading process against interferences from some malicious nodes and be used for the payment transferring of spectrum trading. The remaining deposit can be returned back to the account of corresponding OP. The \textit{LogRegisterOp} event is defined in the \textit{BidOrAskSubmit( )}. If OPs successfully register in the MOSS smart contract and submit their bid, the \textit{LogRegisterOp} event will be triggered. All blockchain nodes listening for this event will know the bid information of registered OPs, from which unregistered OPs can dynamically adjust their demands and bids. The bid information is sent as a transaction via the anonymous address of each OP. Therefore, the real identity of OPs will not be revealed and the private information of OPs will be protected.

\textbf{\emph{RegistrationEnd( )}}: Multi-OPs can call the \textit{RegistrationEnd( )} function at any time to determine whether the registration stage is ended. The timestamp of transactions sent by OPs is represented by now. If now  $>\left( {{\text{t}}_{\text{0}}}+{{\text{t}}_{\text{bid}}} \right)$, the registration stage is already ended. Therefore OPs cannot call the \textit{BidOrAskSubmit( )} function to submit the bid and cannot participate in this round of spectrum sharing.

\subsection{The spectrum trading}

The process of spectrum trading is divided into two sub-stages. The first sub-stage is the spectrum auction among multi-OPs, which mainly includes four functions: the \textit{sortAskByIncrease( )}, the \textit{sortBidByDecrease( )}, the \textit{DoubleAuction( )} and the \textit{doubleAuctionFinish( )}. The second sub-stage is mainly designed for OPs who fail to match in the stage of spectrum auction and provides a free-trading market for unsuccessfully matched OPs. The second sub-stage includes four functions: the \textit{freeTradeBegin( )}, the \textit{orderResponse( )}, the \textit{deleteOrder( )}, and the \textit{MarketEnd( )}. The specific functionalities of functions contained in two sub-stages are described as follows.

\textbf{\emph{sortAskByIncrease( )}}: As the registration stage of OPs ended, the administrator calls the \textit{sortAskByIncrease( )} in the MOSS smart contract to sort $\text{M}$ seller OPs by the price in ascending order.

\textbf{\emph{sortBidByDecrease( )}}: The administrator also calls the \textit{sortBidByDecrease( )} to sort $\text{N}$ buyer OPs by the price in descending order.

\begin{algorithm}[!h]
\caption{Spectrum Auction}
\LinesNumbered 
\KwIn{\textit{asks}=$\left\{\left({{p}_{1}},{{B}_{1}},addr_{1}^{S}\right),\cdots,\left({{p}_{M}},{{B}_{M}},addr_{M}^{S} \right) \right\}$ \newline // $\left( {{p}_{1}}\le \cdots \le {{p}_{M}} \right)$ \newline  \textit{bids}=$\left\{ \left( {{c}_{1}},{{W}_{1}},addr_{1}^{B} \right),\cdots ,\left( {{p}_{N}},{{W}_{N}},addr_{N}^{B} \right) \right\}$ \newline // $\left( {{c}_{1}}\ge \cdots \ge {{c}_{N}} \right)$ \newline \emph{doubleAuctionFinish}=false}
\KwOut{ Successfully matched result}
\textit{bidsLength}=\emph{bids}.length;

\textit{asksLength}=\emph{asks}.length;

\textit{doubleAuctionFinish}=false;

\While{ $bidsLength!=0 \And asksLength!=0 \And \left( {{c}_{1}}\ge {{p}_{1}} \right)$ }{

\textit{dealPrice}=$\left( {{p}_{1}}+{{c}_{1}} \right)/2$;

\textit{dealAmount}=$\min \left\{ {{B}_{1}},{{W}_{1}} \right\}$;

\textit{totalMoney}=\textit{dealPrice}*\textit{dealAmount};

$\text{deposit}\left[ addr_{1}^{B} \right]=\text{deposit}\left[ addr_{1}^{B} \right]-totalMoney$;

$\text{deposit}\left[ addr_{1}^{S} \right]=\text{deposit}\left[ addr_{1}^{S} \right]+totalMoney$;

$\left( {{p}_{1}},{{B}_{1}} \right)\leftarrow \left( {{p}_{1}},{{B}_{1}}-dealAmount \right)$;

$\left( {{c}_{1}},{{W}_{1}} \right)\leftarrow \left( {{c}_{1}},{{W}_{1}}-dealAmount \right)$;

\eIf{$\left( {{W}_{1}}\text{==}0 \right)$}{
\For{$\left( {{c}_{n}},{{W}_{n}},addr_{n}^{B} \right)\in bids$}{
$\left( {{c}_{n}},{{W}_{n}},addr_{n}^{B} \right)\leftarrow \newline
                        \indent \indent \left( {{c}_{n+1}},{{W}_{n+1}},addr_{n+1}^{B} \right)$}
}{\If {$\left( {{B}_{1}}\text{==}0 \right)$}{
\For{$\left( {{p}_{m}},{{B}_{m}},addr_{m}^{S} \right)\in asks$}{
$\left( {{p}_{m}},{{B}_{m}},addr_{m}^{S} \right)\leftarrow \newline
                        \indent \indent\left( {{p}_{m+1}},{{B}_{m+1}},addr_{m+1}^{S} \right)$
}
}}
}
\textit{doubleAuctionFinish}=true
\end{algorithm}

\textbf{\emph{DoubleAuction( )}}: The administrator calls the \textit{DoubleAuction( )} to auction the spectrum for all registered OPs after the sorting is completed. The process of \textit{DoubleAuction( )} is shown in Algorithm 1. The consensus nodes who packages the specific transaction into the block performs the corresponding operations:
\begin{enumerate}
\item  registered buyer OPs and seller OPs are matched based on the spectrum auction algorithm defined in the \textit{DoubleAuction( )}. The successfully matched price per unit bandwidth is set as the average price of the matched buyer OP and seller OP.
\item Second, based on the matched price and bandwidth quantity, the expense of buyer OPs and the income of seller OPs are calculated.
\item Third, the spectrum auction is finished until the bidding queue have been fully matched or the lowest seller price is higher than the highest buyer price.
\end{enumerate}
 The variable \textit{doubleAuctionFinish} defined in the MOSS smart contract represents the state of spectrum auction. When the value of \textit{doubleAuctionFinish} changes from false to true, all OPs and the administrator who inquire the \textit{doubleAuctionFinish} can know that the spectrum allocation has already completed. The \textit{LogDealRecord} event is defined in the MOSS smart contract. When buyer OPs and seller OPs have a successful match, the \textit{LogDealRecord} will be triggered. All nodes listening for the \textit{LogDealRecord} will know the successful match information. The repeated auction of spectrum and the mutual interferences of spectrum usage are avoided based on the proposed spectrum allocation scheme. The openness of spectrum sharing can be guaranteed in the proposed solution.

\textbf{\emph{freeTradeBegin( )}}: If there exist unsuccessful matched OPs in the \textit{DoubleAuction( )}, the administrator will call the \textit{freeTradeBegin( )} at time ${{t}_{1}}$ to open the free-trading market. The valid period of free-trading market is $\left[ {{\text{t}}_{\text{1}}}\text{ , }{{\text{t}}_{\text{1}}}+{{\text{t}}_{\text{free}}} \right]$.

\textbf{\emph{orderResponse( )}}: During the period of $\left[ {{\text{t}}_{\text{1}}}\text{ ,  }{{\text{t}}_{\text{1}}}+{{\text{t}}_{\text{free}}} \right]$, OPs who fail to match in the \textit{DoubleAuction( )} can choose whether to enter the free-trading market. Based on the matched information in the spectrum auction stage, seller OPs who decide to enter the free-trading market can call the \textit{orderResponse( )} to resubmit the adjusted price and the bandwidth demand. Once the latest bid information is submitted successfully, the \textit{LogFreeMarketOrder} event is triggered to declare the newly updated price information of OPs. Buyer OPs can call the \textit{orderResponse( )} to purchase the desired spectrum based on the latest spectrum information and reach a successful match. All the updated bid information and successful match result will be recorded in the permissioned blockchain by consensus nodes. The flow of \textit{orderResponse( )} is shown in Algorithm 2.

\textbf{\emph{deleteOrder( )}}: If the OP fails to match in the \textit{DoubleAuction( )} and refuses to enter the free-trading market, the OP can call the \textit{deleteOrder( )} to clear the bid information and exit the spectrum sharing market.

\textbf{\emph{MarketEnd( )}}: Multi-OPs and the administrator can call the \textit{MarketEnd( )} to determine whether the free-trading market is ended. If now $>\left( {{\text{t}}_{\text{1}}}+{{\text{t}}_{\text{free}}} \right)$, the OP cannot call the \textit{orderResponse( )} since the free-trading market has already ended. When the free-trading market is ended, the whole spectrum sharing process is finished. If there exist OPs who still want to join the spectrum sharing, they can only wait for the next new round of spectrum sharing.

\begin{algorithm}[!h]
\caption{Free-trading Market}
\LinesNumbered 
\KwIn{$\text{ }\!\!\_\!\!$ {\textit{ addr}} , $\text{ }\!\!\_\!\!$ \textit{ price} , $\text{ }\!\!\_\!\!$ \textit{bandwidth}}
\KwOut{Successfully matched result}
\If{$\text{Buyer OPs purchase the spectrum}$}{
\textit{new }$\!\!\_\!\!$ \textit{ price= }$\!\!\_\!\!$ \textit{ price};

\textit{new }$\!\!\_\!\!$ \textit{ band= }$\!\!\_\!\!$ \textit{ bandwidth};
}

\If{OPs resubmit the adjusted bids}{
\eIf{$\left( \_ \textit{bandwidth}\le \text{ }\!\!\_ \textit{  addr}.\emph{bandwidth} \right)$}{
$totalMoney= \!\!\_ \textit{  price}* \!\!\_  \textit{ bandwidth}$;\\
$\text{deposit}\left[ msg{.sender} \right] \leftarrow \newline
\indent \indent \text{deposit}\left[ msg{.sender} \right]{-totalMoney}$;\\
$\text{deposit}\left[ \text{ }\!\!\_\textit{ addr} \right] \leftarrow \newline
\indent \indent \text{deposit}\left[ \text{ }\!\!\_\textit{ addr} \right]{+totalMoney}$;\\
${msg}{.sender}{.bandwidth=0}$;\\
$\text{ }\!\!\_\textit{ addr}\textit{.bandwidth= }\_\textit{ addr}\emph{.bandwidth- }\_\textit{ bandwidth}$;\\
}{
\If{$\left( \text{ }\!\!\_\textit{ bandwidth }> \text{ }\!\!\_ \textit{  addr}.\emph{bandwidth} \right)$}{
$\textit{totalMoney=  }\_ \textit{ price* }\_\textit{ addr}{.bandwidth}$;\\
$\text{deposit}\left[ {msg}{.sender} \right]\leftarrow \newline
                         \indent\indent \text{deposit}\left[ {msg}{.sender} \right]{-totalMoney}$;\\
$\text{deposit}\left[ \text{ }\!\!\_\textit{ addr} \right] \leftarrow \newline
                         \indent \indent \text{deposit}\left[ \text{ }\!\!\_\textit{ addr} \right]{+totalMoney}$;\\
${msg}{.sender}{.bandwidth} \leftarrow \newline
                         \indent \indent {msg}{.sender}\textit{.bandwidth- }\_\textit{ bandwidth}$;\\
$\text{ }\!\!\_\textit{ addr}{.bandwidth=0}$;
}
}
}

\end{algorithm}

\subsection{The payment clearing}
In order to implement the punishment of malicious OPs and the payment transferring of registered OPs, three functions including the \textit{payORnot ( )}, the \textit{increaseFunds( )} and the \textit{withdraw( )} are designed in the MOSS smart contract. The specific functionalities of \textit{payORnot ( )}, \textit{increaseFunds( )} and \textit{withdraw( )} are described as follows.

\textbf{\emph{payORnot( )}}: A function modifier \textit{ownerOnly} is set in the \textit{payORnot ( )}, which regulates that the \textit{payORnot ( )} can only be called by the administrator. Other blockchain nodes do not have the permission to call the \textit{payORnot ( )}. During the period of $\left[ {{\text{t}}_{\text{b}}}\text{ , }{{\text{t}}_{\text{e}}} \right]$, the administrator is responsible for ensuring that the exchange of spectrum usage right is executed off-chain correctly. The bool variable \textit{executeORnot} is defined in the MOSS smart contract to represent whether the corresponding OP correctly exchanges the spectrum usage right. If the OP does not exchange the spectrum usage right based on matched results of the MOSS smart contract, the administrator can call the \textit{payORnot ( )} to set the \textit{executeORnot} value of corresponding OP as false. Thus malicious OPs are prevented from withdrawing their remaining deposit. As for honest OPs, the administrator calls the \textit{payORnot ( )} to set the \textit{executeORnot} of honest OPs as true and honest OPs can successfully withdraw their remaining deposit.

\textbf{\emph{increaseFunds( )}}: The \textit{increaseFunds( )} is defined as payable for multi-OPs to increase their deposit in the MOSS smart contract. The \textit{increaseFunds( )} is called by OPs when the remaining deposit is not enough to be transferred to other OPs. Thus the payment transferring of spectrum trading can be executed smoothly.

\textbf{\emph{withdraw( )}}: All OPs participating in the spectrum sharing will call the \emph{withdraw( )} to get back the remaining deposit at the end time of $\left[ {{\text{t}}_{\text{b}}}\text{ , }{{\text{t}}_{\text{e}}} \right]$. When OPs whose value of the \textit{executeORnot} are false call the \textit{withdraw( )} to get back the remaining deposit, the consensus nodes will not package this transaction. Honest Ops whose value of the \textit{executeORnot} are true can smoothly take back the remaining deposit by calling the \textit{withdraw( )}.

The proposed MOSS smart contract can resolve the problem with the existence of malicious OPs in the spectrum sharing. The administrator can punish malicious OPs through the MOSS smart contract, making them unable to take back the deposit. A friendly and secure spectrum sharing market can be built by prompting all OPs honestly exchange the spectrum usage right.

\section{SYSTEM IMPLEMENTATION AND ANALYSIS}
In this senction, the designed MOSS smart contract is implemented and tested in the Remix IDE found at \url{http://remix.ethereum.org}. The Remix is a browser-based IDE for developing smart contracts without the need to install the Solidity environment, which is an object-oriented, high-level language for implementing smart contracts. The performance analysis of the proposed solution are also given in the following.

\subsection{Testing of the MOSS Smart Contract}
In order to test and estimate the cost of MOSS smart contract visually, the designed MOSS smart contract is compiled in the Remix IDE and deployed in the Ganache, which is used for the test case. Nodes can send transactions to deploy the MOSS smart contract and call functions through their unique addresses. The gas price and gas limit can be independently set by OPs when transactions are sent.

Three seller OPs and three buyer OPs are considered to test the feasibility of the MOSS smart contract. The addresses and spectrum demands of all entities are shown in Table 2. Based on accounts provided by the Ganache, the MOSS smart contract is tested and the gas consumption of each function in the MOSS smart contract is estimated. The Ether consumption can be calculated as
\begin{equation}
\begin{split}
  \text{Ether }\!\!\_\!\!\text{ consumption=Gas }\!\!\_\!\!\text{ cos}{{\text{t}}^{{}}}{{\text{*}}^{{}}}\text{Gas }\!\!\_\!\!\text{ price}
   \end{split}.
\label{eq}
\end{equation}

\begin{table}
\caption{Accounts and spectrum bid information of each entity.}
\label{table}
\setlength{\tabcolsep}{3pt}
\begin{tabular}{|c|c|c|c|}
\hline
Role&
Account Address&
Amount(MHZ)&
Price(Gwei/MHZ)\\
\hline
administrator&
0xd769 \ldots e4AC&
-&
- \\
seller ${{\operatorname{OP}}_{1}}$&
0x36d2 \ldots E1fE&
20&
2000000\\
seller ${{\operatorname{OP}}_{2}}$&
0x5E6E \ldots f1A0&
10&
1600000\\
seller ${{\operatorname{OP}}_{3}}$&
0x0004 \ldots 561d&
15&
2400000\\
buyer ${{\operatorname{OP}}_{4}}$&
0x5B84 \ldots 2631&
10&
1500000\\
buyer ${{\operatorname{OP}}_{5}}$&
0x8c66 \ldots 1155&
12&
2500000\\
buyer ${{\operatorname{OP}}_{6}}$&
0x9508 \ldots bcaa&
8&
1800000\\
\hline
\end{tabular}
\label{tab1}
\end{table}

Based on the current recommended gas price found at \url{https://ethgasstation.info/}, the gas price is set as 4.3 Gwei, where wei is the minimum unit of ether and $1{}^{{}}ether={{10}^{9}}Gwei\text{=}{{10}^{18}}wei$. Considering that the gas consumption is related to different operations defined in the functions, all the gas consumption corresponded to the possible function execution results are shown in Table 3. The corresponding Ether cost is also calculated and given in Table 3. From Table 3, the Ether cost of deploying the MOSS smart contract is 0.0204989 \textit{eth}. Among all functions called by the administrator in the MOSS smart contract, the function which costs the most amount of Ether is \textit{DoubleAuction( )} with the cost of 0.001583935 \textit{eth}. The Ether cost of remaining functions are modest.

\begin{table}
\caption{Gas cost of functions called by the Administrator.}
\label{table}
\setlength{\tabcolsep}{4pt}
\begin{tabular}{|p{70pt}|p{80pt}|p{60pt}|}
\hline
Function&
Transaction Gas&
Gas cost in Ether
\\
\hline
MOSS smart contract&
4767204 gas&
${\text{2}}{\text{.04989}} \times {\text{1}}{{\text{0}}^{{\text{ - 2}}}}$ \\
\textit{RegistrationEnd( )}&
21799 gas
&
${\text{9}}{\text{.37357}} \times {\text{1}}{{\text{0}}^{{\text{ - 5}}}}$ \\
\textit{sortAskByIncrease( )}&
70696 gas&
${\text{3}}{\text{.0399}} \times {\text{1}}{{\text{0}}^{{\text{ - 4}}}}$ \\
\textit{sortBidByDecrease( )}&
116557 gas&
${\text{5}}{\text{.01195}} \times {\text{1}}{{\text{0}}^{{\text{ - 4}}}}$ \\
\textit{DoubleAuction( )}&
368357 gas&
${\text{1}}{\text{.583935}} \times {\text{1}}{{\text{0}}^{{\text{ - 3}}}}$ \\
\textit{freeTradeBegin( )}&
42413 gas&
${\text{1}}{\text{.82375}} \times {\text{1}}{{\text{0}}^{{\text{ - 4}}}}$ \\
\textit{MarketEnd( )}&
21776 gas
&
${\text{9}}{\text{.35938}} \times {\text{1}}{{\text{0}}^{{\text{ - 5}}}}$ \\
\textit{payORnot( )}&
29018 gas&
${\text{1}}{\text{.24777}} \times {\text{1}}{{\text{0}}^{{\text{ - 4}}}}$ \\
\textit{changeOwner( )}&
28811 gas&
${\text{1}}{\text{.23887}} \times {\text{1}}{{\text{0}}^{{\text{ - 4}}}}$ \\
\textit{selfDestruct( )}&
13495 gas&
${\text{5}}{\text{.8028}} \times {\text{1}}{{\text{0}}^{{\text{ - 5}}}}$ \\
\hline
\end{tabular}
\label{tab2}
\end{table}

The gas consumption of all functions called by OPs and the cost of Ether are shown in Table 4. From Table 4, it is obvious that only the \textit{BidOrAskSubmit( )}, the cost of which is 0.001 \textit{eth}, consumes the most gas among all functions called by OPs in the MOSS smart contract. And the overall cost of remaining functions called by OPs are modest.

\begin{table}[!h]
\caption{Gas cost of functions called by OPs.}
\label{table}
\setlength{\tabcolsep}{4pt}
\begin{tabular}{|p{70pt}|p{90pt}|p{60pt}|}
\hline
Function&
Transaction Gas&
Gas cost in Ether\\
\hline
\textit{BidOrAskSubmit( )}&
216416 gas &
${\text{1}}{\text{.00362}} \times {\text{1}}{{\text{0}}^{{\text{ - 3}}}}$\\
\textit{deleteOrder( )}&
21229 gas&
${\text{9}}{\text{.12847}} \times {\text{1}}{{\text{0}}^{{\text{ - 5}}}}$ \\
\textit{orderResponse( )}&
24277 gas when buyer OPs call this function;\par
35085 gas when seller OPs want to change the bid&
${\text{1}}{\text{.0439}} \times {\text{1}}{{\text{0}}^{{\text{ - 4}}}}$  $\sim$ \par ${\text{1}}{\text{.5086}} \times {\text{1}}{{\text{0}}^{{\text{ - 4}}}}$ \\
\textit{withdraw( )}&
22188 gas &
${\text{9}}{\text{.5408}} \times {\text{1}}{{\text{0}}^{{\text{ - 5}}}}$ \\
\textit{increaseFunds( )}&
26757 gas&
${\text{1}}{\text{.15055}} \times {\text{1}}{{\text{0}}^{{\text{ - 4}}}}$ \\
\hline
\end{tabular}
\label{tab3}
\end{table}

From Table 3 and Table 4, the gas cost called by OPs are almost negligible compared to the cost of functions called by the administrator. OPs can trade the spectrum with the modest cost, which greatly motivates OPs to participate in the proposed spectrum sharing market and improves the spectrum utilization. Different from the spectrum broker that exists as a third party in previous studies, the administrator does not directly participate in the spectrum allocation among multi-OPs in our proposed solution. The spectrum allocation among multi-OPs is implemented based on predefined functions in the MOSS smart contract. The administrator only acts as a deployer and caller of functions in the MOSS smart contract, and supervises the entire spectrum sharing process off-chain.

\begin{figure}[!h]
\centering
\subfloat[Transaction log]{\includegraphics[width=0.48\textwidth]{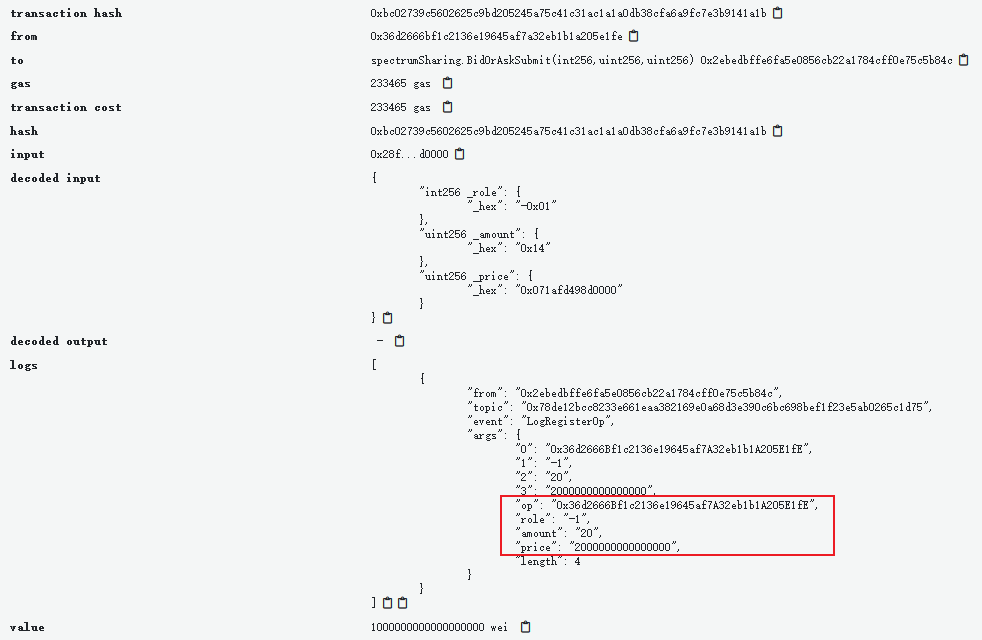}%
\label{fig_Transaction log}}\\
\centering
\subfloat[Account balance]{\includegraphics[width=0.3\textwidth]{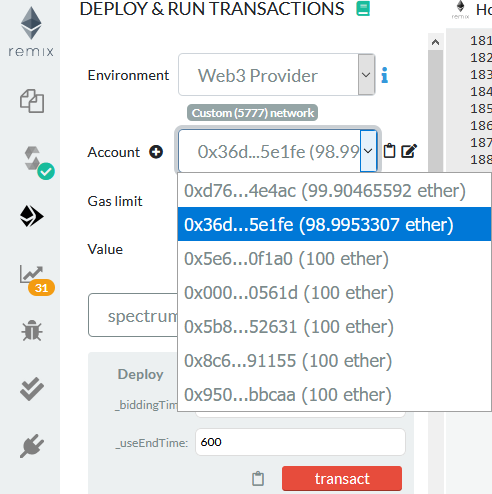}%
\label{fig_Account Balance }}
\caption{Calling the \textit{BidOrAskSubmit( )} by ${{\operatorname{OP}}_{1}}$.}
\label{fig3}
\vspace{-0.3cm}
\end{figure}

The \textit{LogRegisterOp} event is triggered when OPs call the \textit{BidOrAskSubmit( )}. Figure 3(a) shows the transaction log when the seller ${{\operatorname{OP}}_{1}}$ calls the \textit{BidOrAskSubmit( )} to submit the bid information. From the log field shown in Figure 3(a), the address of ${{\operatorname{OP}}_{1}}$ is 0x36d...E1Fe, the amount of bandwidth that the ${{\operatorname{OP}}_{1}}$ wants to sell is 20MHZ in the simulation and the price per unit bandwidth is set as 2000000 \textit{Gwei}. All nodes in the blockchain that listen to the \textit{LogRegisterOp} will receive the latest registration information of ${{\operatorname{OP}}_{1}}$. Other OPs can reasonably adjust their own bids based on the bid information of ${{\operatorname{OP}}_{1}}$. Meanwhile, OPs can set their own price in the next round of spectrum sharing by learning the bid information of other OPs in each round. The probability of successful match in the spectrum auction process will increase. The account balance of ${{\operatorname{OP}}_{1}}$ is almost 98.99 \textit{eth} shown in Figure 3(b). The ${{\operatorname{OP}}_{1}}$ is required to deposit 1 \textit{eth} in the MOSS smart contract when the \textit{BidOrAskSubmit( )} is called. The gap between the account balance of ${{\operatorname{OP}}_{1}}$ and 99 \textit{eth} is the transaction fee when the transaction is sent to call the \textit{BidOrAskSubmit( )}.

\begin{figure}[h]
\centerline{\includegraphics[width=\columnwidth]{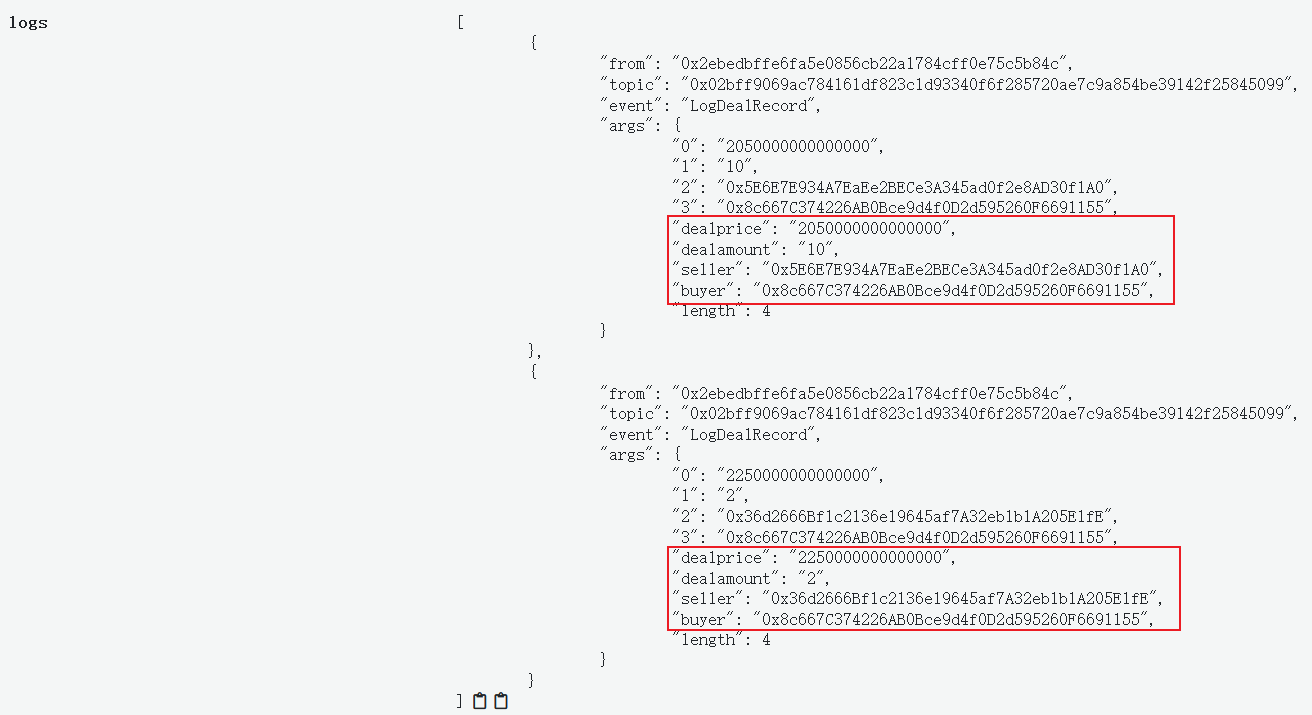}}
\caption{Transaction log of calling the \textit{DoubleAuction( )} by the Administrator.}
\label{fig4}
\end{figure}

Figure 4 shows the transaction log when the administrator calls the \textit{DoubleAuction( )}. The predefined spectrum auction algorithm in the \textit{DoubleAuction( )} is executed by consensus nodes. If there exist successfully matches among multi-OPs in the spectrum auction, the \textit{LogDealRecord} is triggered. It can be seen from the log field in Figure 4 that the seller ${{\operatorname{OP}}_{2}}$ and the buyer ${{\operatorname{OP}}_{5}}$ have successfully traded 10 MHz of spectrum at a price of 2050000 \textit{Gwei}. Also, the seller ${{\operatorname{OP}}_{1}}$ and the buyer ${{\operatorname{OP}}_{5}}$ have successfully traded 2 MHz of spectrum at a price of 2250000 \textit{Gwei}. The remaining OPs fail to match in the \textit{DoubleAuction( )} stage.

After the spectrum auction, the \textit{freeTradeBegin( )} is called by the administrator to open the spectrum free-trading market among unsuccessfully matched OPs. In the simulation, the seller ${{\operatorname{OP}}_{1}}$ is assumed to enter the free-trading market and modify the bid price to 1800000 \textit{Gwei} by calling the \textit{orderResponse( )}. The remaining unsuccessfully matched buyer OPs can decide whether to participate in the free-trading market based on the latest bid information of ${{\operatorname{OP}}_{1}}$. If the buyer ${{\operatorname{OP}}_{6}}$ decides to purchase the spectrum after listening to the modified bid information of ${{\operatorname{OP}}_{1}}$, the buyer ${{\operatorname{OP}}_{6}}$ can call the \textit{orderResponse( )} to declare the desired spectrum information and reach a successful match. It can be seen from the transaction log field in Figure 5 that the buyer ${{\operatorname{OP}}_{6}}$ and the seller ${{\operatorname{OP}}_{1}}$ have successfully traded 8 MHz spectrum at a price of 1800000 \textit{Gwei}.

\begin{figure}[h]
\centerline{\includegraphics[width=\columnwidth]{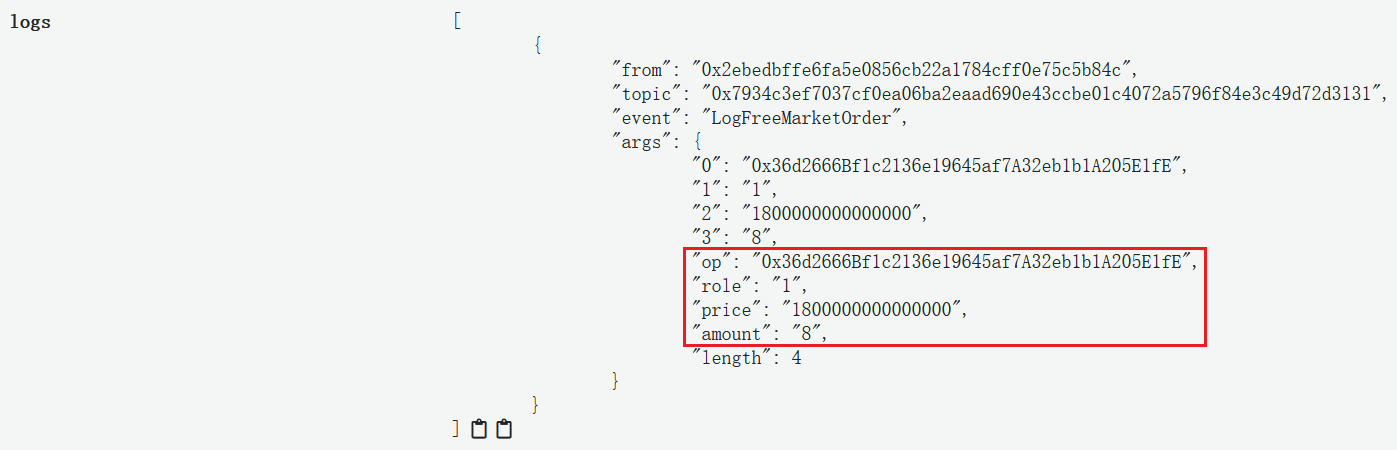}}
\caption{Transaction log of calling the \textit{orderResponse( )} by the buyer ${{\operatorname{OP}}_{6}}$.}
\label{fig5}
\end{figure}

\begin{figure}[h]
\centerline{\includegraphics[width=\columnwidth]{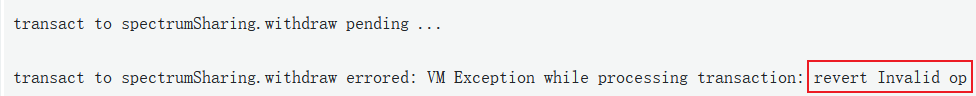}}
\caption{Transaction log of calling the \textit{withdraw( )} by the seller ${{\operatorname{OP}}_{2}}$.}
\label{fig6}
\end{figure}

When the free-trading market is ended, the entire spectrum sharing process is completed. Each OP can call the \textit{withdraw( )} to withdraw the remaining deposit in the MOSS smart contract. The seller ${{\operatorname{OP}}_{2}}$ is assumed to be a malicious OP in the simulation. After successfully matched with the buyer ${{\operatorname{OP}}_{5}}$ in the previous spectrum auction, the traded spectrum usage right of ${{\operatorname{OP}}_{2}}$ is not correctly delivered to the corresponding buyer ${{\operatorname{OP}}_{5}}$ within the period of $\left[ {{\text{t}}_{\text{b}}}\text{ , }{{\text{t}}_{\text{e}}} \right]$. Once supervising the off-chain malicious behaviour of ${{\operatorname{OP}}_{2}}$ , the administrator can call the \textit{payORnot( )} to set the \textit{executeORnot} of ${{\operatorname{OP}}_{2}}$ as false and freeze the account balance of ${{\operatorname{OP}}_{2}}$. If the \textit{withdraw( )} function is called to withdraw the deposit by the seller ${{\operatorname{OP}}_{2}}$, the transaction is rolled back and declared as "Invalid op" shown in Figure 6. Other honest OPs can successfully withdraw their deposits by calling the \textit{withdraw( )}. It can be seen from Figure 7(a) that the transaction log field of honest ${{\operatorname{OP}}_{1}}$ is shown as "successful to withdraw". The account balance of ${{\operatorname{OP}}_{1}}$ is more than 100 \textit{eth} shown in Figure 7(b), which indicates that the seller ${{\operatorname{OP}}_{1}}$ can withdraw its 1 \textit{eth} deposit and earn the extra revenue from selling the spectrum usage right.

\begin{figure}[!h]
\centering
\subfloat[Transaction log]{\includegraphics[width=0.48\textwidth]{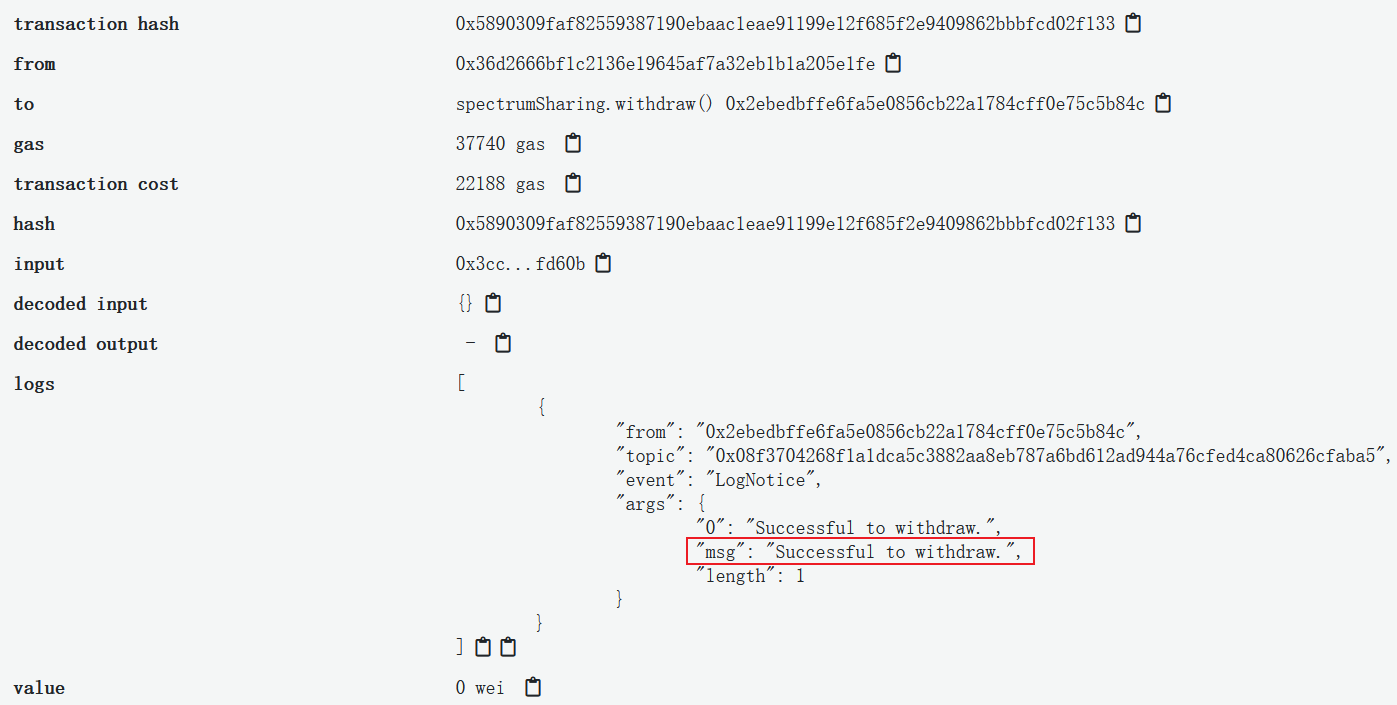}%
\label{fig_Transaction log}}\\
\centering
\subfloat[Account balance]{\includegraphics[width=0.3\textwidth]{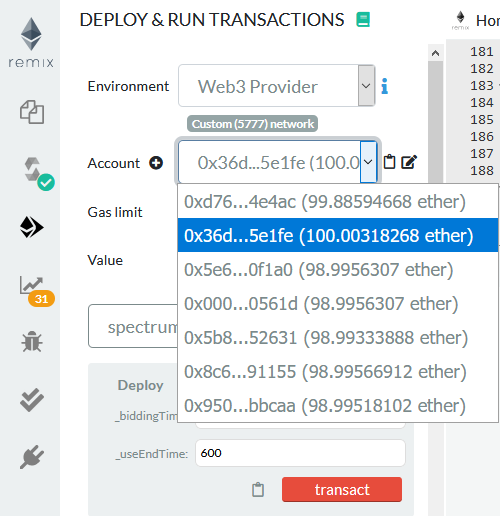}%
\label{Accout Balance}}
\caption{Calling the \textit{withdraw( )} by ${{\operatorname{OP}}_{1}}$.}
\label{fig7}
\vspace{-0.3cm}
\end{figure}

\subsection{Performance analysis}

\subsubsection{SECURITY ANALYSIS}
In traditional multi-OPs spectrum sharing solutions, the spectrum broker is introduced to allocate and manage spectrum resources. The problem of privacy disclosure exists when multi-OPs interact with the spectrum broker. Besides, the data sent by multi-OPs to the trustless spectrum broker may be maliciously tampered with. In this paper, the distributed chained-block structure of blockchain technology is used to realize the spectrum sharing among multi-OPs. All transactions and spectrum usage status are recorded in a data structure of merkle tree in the blockchain. All transactions need to be signed with the private key of transaction sender and verified by the public key of transaction receiver. The private key and the public key of nodes are generated by the asymmetric encryption algorithm. Each consensus node in the blockchain locally maintains a distributed ledger to record all transactions broadcasted in the blockchain. If there exist malicious OPs who want to tamper with the transaction information, honest OPs who account for a large proportion will not modify their own local ledger. The integrity and security of data can be ensured by the proposed solution. Besides, the permissioned blockchain among multi-OPs is constructed in this paper. Different from the public blockchain, the private data of OPs is invisible to nodes outside the permissioned blockchain. Moreover, through the certificate authentication (CA) of permissioned blockchain, strict identity management is conducted for all OPs participating in the spectrum sharing. The privacy preserving of multi-OPs in the spectrum sharing process can be realized in the proposed solution.
\subsubsection{RELIABILITY ANALYSIS}
Based on the decentralization characteristic of blockchain, each OP in the permissioned blockchain maintains a local ledger in a distributed structure. When one of OPs fails, other OPs can still normally operate. The single point of failure is avoided in the proposed solution, which is more reliable than traditional centralized spectrum sharing solutions. Besides, the MOSS smart contract is designed to implement the spectrum allocation among multi-OPs, eliminating the existence of centralized spectrum broker in traditional methods. Before starting the spectrum sharing, multi-OPs jointly design the logic of spectrum allocation scheme in the MOSS smart contract. The entire spectrum sharing process is automatically executed based on the agreed functions in the MOSS smart contract, which ensures the fairness and openness of spectrum sharing in multi-OPs wireless communication networks.
\begin{table}[!h]
\centering
\caption{Comparison of proposed framework with existing researches.}
\label{table}
\setlength{\tabcolsep}{3pt}
\begin{tabular}{|c|c|c|c|c|c|}
\hline
Characteristic &[7] & [8]  & [15] &[25]& Proposed solution \\\hline
Decentralized & $\times$ & $\surd$  & $\surd$ & $\surd$ & $\surd$\\
Privacy & $\times$ & $\times$ & $\times$ & $\surd$ & $\surd$   \\
Traceable & $\times$ & $\times$  & $\surd$ &$\surd$&$\surd$ \\
MOSS &$\times$ &$\times$  & $\times$ &$\times$ &$\surd$ \\
fairness & $\surd$ & $\surd$ & $\surd$ &$\surd$& $\surd$ \\
Openness & $\times$ & $\times$ & $\times$ & $\surd$ & $\surd$\\
\hline
\end{tabular}
\label{tab1}
\end{table}
\subsubsection{COMPARISON ANALYSIS}
In order to have an intuitive analysis, our proposed framework is compared to other three existing researches of the spectrum sharing among multi-OPs. The comparison results are shown in Table 5. The analysis results demonstrate that the privacy, openness and fairness of the proposed solution are better than other three spectrum sharing solutions.

\section{Conclusion}
In this paper, a permissioned blockchain trust framework is proposed for the spectrum sharing in multi-OPs wireless communication networks. The MOSS smart contract is designed to automatically perform the spectrum trading and payment transferring among multi-OPs. A fair spectrum sharing market among multi-OPs can be realized without the existence of third party. The gas consumption of each function defined in the MOSS smart contract is tested. The security and the reliability analysis of the proposed solution are evaluated. The simulation results and the performance analysis demonstrate that the privacy, openness and fairness of the proposed solution are better than traditional spectrum allocation solutions.

In the future work, the incentive mechanism will be considered to incentive OPs to share their spectrum, which can further improve the spectrum efficiency. Besides, designing an efficient auction algorithm will be considered to further reduce the gas cost of calling functions in the MOSS smart contract.

\begin{IEEEbiography}[{\includegraphics[width=1in,height=1.25in,clip,keepaspectratio]{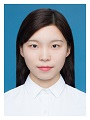}}]{Shuang Zheng}
received the B.E. degree in electronic information engineering from the HuaZhong university of science and technology, Wuhan, China, in 2018. She is currently working toward the Master's degree with the School of Electronic Information and Communications, Huazhong University of Science and Technology. Her research interests are Spectrum sharing and blockchain technology.
\end{IEEEbiography}

\begin{IEEEbiography}[{\includegraphics[width=1in,height=1.25in,clip,keepaspectratio]{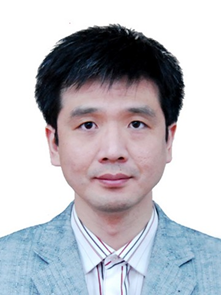}}]{Tao Han}
(M'13) received the Ph.D. degree in information and communication engineering from Huazhong University of Science and Technology (HUST), Wuhan, China in December, 2001. He is currently an Associate Professor with the School of Electronic Information and Communications, HUST. From August, 2010 to August, 2011, he was a Visiting Scholar with University of Florida, Gainesville, FL, USA, as a Courtesy Associate Professor. His research interests include wireless communications, multimedia communications, and computer networks. Dr. Han is currently serving as an Area Editor for the EAI Endorsed Transactions on Cognitive Communications.
\end{IEEEbiography}

\begin{IEEEbiography}[{\includegraphics[width=1in,height=1.25in,clip,keepaspectratio]{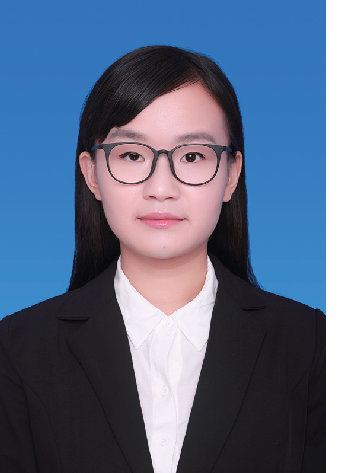}}]{Yuna Jiang}
received the B.E. degree in communications engineering from the China University of Mining and Technology, Xuzhou, China, in 2017. She is currently working toward the Ph.D. degree with the School of Electronic Information and Communications, Huazhong University of Science and Technology. Her research interests are Internet of Things and blockchain technology.
\end{IEEEbiography}

\begin{IEEEbiography}[{\includegraphics[width=1in,height=1.25in,clip,keepaspectratio]{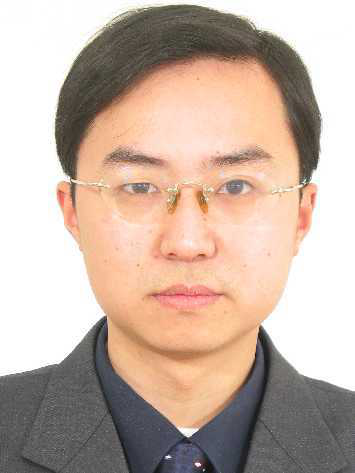}}]{Xiaohu Ge}
(M'09--SM'11) is currently a full Professor with the School of Electronic Information and Communications at Huazhong University of Science and Technology (HUST), China. He is an adjunct professor with the Faculty of Engineering and Information Technology at University of Technology Sydney (UTS), Australia. He received his PhD degree in Communication and Information Engineering from HUST in 2003. He has worked at HUST since Nov. 2005. Prior to that, he worked as a researcher at Ajou University (Korea) and Politecnico Di Torino (Italy) from Jan. 2004 to Oct. 2005. His research interests are in the area of mobile communications, traffic modeling in wireless networks, green communications, and interference modeling in wireless communications. He has published about 200 papers in refereed journals and conference proceedings and has been granted about 25 patents in China. He services as an IEEE Distinguished Lecturer and an Associate Editor for the IEEE ACCESS, IEEE Wireless Communications and IEEE Transactions on Vehicular Technology.

\end{IEEEbiography}

%

%
%
%




\end{document}